\documentclass[journal]{IEEEtran}

\usepackage{amsmath,bm,amssymb,amsfonts,amsthm}
\usepackage{algorithm,algorithmic}
\usepackage{graphicx}
\usepackage{xcolor}
\usepackage{relsize}
\usepackage{textcomp}
\usepackage{listings}

\usepackage{changepage}
\usepackage{hhline}
\usepackage{multirow}
\usepackage{nomencl}
\usepackage{mathtools}
\usepackage{commath}
\usepackage{booktabs}
\usepackage{subfiles}
\usepackage{tabularx}
\usepackage{lscape}
\usepackage{rotating}

\usepackage[normalem]{ulem}
\usepackage[utf8]{inputenc}
\usepackage[hyphens]{url}
\usepackage{lineno}
    \modulolinenumbers[5]
\usepackage[caption=false]{subfig}
\usepackage{enumitem}
    \setlist{nolistsep}
\usepackage{gensymb}

\usepackage[noadjust]{cite}

\usepackage{scalerel}
\usepackage{tikz}
\usetikzlibrary{svg.path}
\definecolor{orcidlogocol}{HTML}{A6CE39}
\tikzset{
  orcidlogo/.pic={
    \fill[orcidlogocol] svg{M256,128c0,70.7-57.3,128-128,128C57.3,256,0,198.7,0,128C0,57.3,57.3,0,128,0C198.7,0,256,57.3,256,128z};
    \fill[white] svg{M86.3,186.2H70.9V79.1h15.4v48.4V186.2z}
                 svg{M108.9,79.1h41.6c39.6,0,57,28.3,57,53.6c0,27.5-21.5,53.6-56.8,53.6h-41.8V79.1z M124.3,172.4h24.5c34.9,0,42.9-26.5,42.9-39.7c0-21.5-13.7-39.7-43.7-39.7h-23.7V172.4z}
                 svg{M88.7,56.8c0,5.5-4.5,10.1-10.1,10.1c-5.6,0-10.1-4.6-10.1-10.1c0-5.6,4.5-10.1,10.1-10.1C84.2,46.7,88.7,51.3,88.7,56.8z};
  }
}
\newcommand\orcidicon[1]{\href{https://orcid.org/#1}{\mbox{\scalerel*{
\begin{tikzpicture}[yscale=-1,transform shape]
\pic{orcidlogo};
\end{tikzpicture}
}{|}}}}
\usepackage{hyperref}

\makenomenclature

\renewcommand\nomgroup[1]{%
    \ifstrequal{#1}{A}{\item[\textbf{Sets}]}{%
    \ifstrequal{#1}{B}{\vspace{0.10in} \item[\textbf{Parameters}]}{%
    \ifstrequal{#1}{D}{\vspace{0.10in} \item[\textbf{Variables}]}{}}}
    \hspace*{-\leftmargin}\vspace{0.025in}%
}

\nomenclature[A,01]{$E^a$}{set of directed edges in the ac circuit, indexed by $e$}
\nomenclature[A,02]{$E^d$}{set of directed edges in the dc circuit, indexed by $e$ and oriented from dc bus $i$ to dc bus $j$}
\nomenclature[A,03]{$E^{y} \subseteq E^a$}{set of gwye-gwye transformers in the ac circuit, indexed by $e$}
\nomenclature[A,04]{$E^{\Delta} \subseteq E^a$}{set of delta-gwye transformers in the ac circuit, indexed by $e$}
\nomenclature[A,05{$E_i^+ \subseteq E^a \cup E^d$}{set of directed edges connected to node $i$, indexed by $e$, that are oriented from $i$ to $j$}
\nomenclature[A,06]{$E_i^- \subseteq E^a \cup E^d$}{set of directed edges connected to node $i$, indexed by $e$, that are oriented from $j$ to $i$}
\nomenclature[A,07]{$E^{\infty} \subseteq E^a$}{set of auto gwye-gwye transformers in the ac circuit, indexed by $e$ and oriented from bus $i$ to bus $j$}
\nomenclature[A,08]{$G$}{set of generators in the ac circuit, indexed by $g$}
\nomenclature[A,09]{$G_i$}{set of generators in the ac circuit connected to bus $i \in N^a$, indexed by $g$}
\nomenclature[A,10]{$N^{a}$}{set of nodes (buses) in the ac circuit, indexed by $i$}
\nomenclature[A,11]{$N^{d}$}{set of nodes (buses) in the dc circuit, indexed by $i$}
\nomenclature[A,12]{$N^a_l \subseteq N^a$}{set of nodes with ac load}
\nomenclature[A,13]{$T$}{set of time periods, indexed by $t$, numbered from $0\ldots \abs{T}$}

\nomenclature[B,01]{\color{black}$a_{e}$}{dc admittance of edge ${ e }\in {E^d} $}
\nomenclature[B,02]{$a_{i}$}{admittance of the grounding line at dc bus ${i} \in N^d$, 0 if the bus is not the substation neutral}
\nomenclature[B,03]{$e^H, e^L, e^S, e^C \in E^d$}{high side, low side, series, and common winding dc edges for $e \in E^\Delta \cup E^y \cup E^\infty$, respectively}
\nomenclature[B,04]{$\overrightarrow{e}$}{ac edge associated with dc edge $e \in E^d$}
\nomenclature[B,05]{$ g_{i} $}{shunt conductance at bus $ i \in {N^a} $}
\nomenclature[B,06]{$\overline{I}_{e}^d $}{maximum dc current on line $ e \in {E^a} $}
\nomenclature[B,07]{$L_N$,$L_E$}{the north and east components of the displacement of each transmission line, respectively}
\nomenclature[B,08]{$n_e$}{top-oil cooling factor for $e \in E^\Delta \cup E^y \cup E^\infty$ }
\nomenclature[B,09]{$p_i^t $}{real power demand at bus $i \in {N^a}$ at time $t \in T$}
\nomenclature[B,10]{$ \underline{p}_{g} $}{minimum real power generation at generator $g \in G$}
\nomenclature[B,11]{$ \overline{p}_{g} $}{maximum real power generation at generator $i \in G$}
\nomenclature[B,12]{$R_e$}{thermal hot-spot resistance for $e \in E^\Delta \cup E^y \cup E^\infty$ }
\nomenclature[B,13]{$ \overline{s}_{e} $}{apparent power limit on line $e \in {E^a} $}
\nomenclature[B,14]{$\mathcal{V}_e^t$}{GMD induced voltage source at time $t \in T$}
\nomenclature[B,15]{$x_{e} $}{reactance of line $e \in {E^a} $}
\nomenclature[B,16]{$\alpha_e$}{turns ratio for $e \in E^\Delta \cup E^y \cup E^\infty$}
\nomenclature[B,17]{$\delta_e^r$}{top-oil temperature rise in $\degree\mathrm{C}$ at rated apparent power for $e \in E^\Delta \cup E^y \cup E^\infty$}
\nomenclature[B,18]{$\Delta$}{duration of a time period}
\nomenclature[B,19]{$\mathcal{E}_N$,$\mathcal{E}_E$}{strength of the north and east geo-electric field at time $t \in T$, respectively}
\nomenclature[B,20]{$\zeta_e$}{discretized top-oil temperature time constant of $e \in E^\Delta \cup E^y \cup E^\infty$}
\nomenclature[B,21]{$ \theta^M$}{big-M parameter given by $\lvert{E^a}\rvert\overline{\theta}$}
\nomenclature[B,22]{$ \overline{\theta}$}{phase angle difference limit}
\nomenclature[B,23]{$\kappa_{g}^0,\kappa_{g}^1,\kappa_{g}^2$}{generation cost coefficients of generator $g \in G$}
\nomenclature[B,24]{$\rho_e^t$}{ambient temperature at transformer edge $e \in E^\Delta \cup E^y \cup E^\infty$ at time $t \in T$}
\nomenclature[B,25]{$\tau_e$}{top-oil temperature time constant of $e \in E^\Delta \cup E^y \cup E^\infty$}
\nomenclature[B,26]{$\mathcal{T}_e$}{maximum temperature of $e \in E^\Delta \cup E^y \cup E^\infty$}
\nomenclature[B,27]{$\phi^t$}{the angle of the geo-electric field relative to east at time $t \in T$}

\nomenclature[D,01]{$ I_{e}^{t}$}{GIC flow on dc line $e \in E^{d} $ at time $t \in T$}
\nomenclature[D,02]{$ \widetilde{I}_{e}^t$}{the effective GIC on line $e \in E^{a} $  at time $t \in T$}
\nomenclature[D,03]{$ p_{e,ij}^t$}{real power flow on line $ e \in {E^a}$, as measured at node $i$ at time $t \in T$}
\nomenclature[D,04]{$ p_{e,ji}^t$}{real power flow on line $ e \in {E^a}$, as measured at node $j$ at time $t \in T$}
\nomenclature[D,05]{$ p_g^t$}{real and reactive power generated at generator $g \in G$ at time $t \in T$}
\nomenclature[D,06]{$ q_{e,ij}^t$}{reactive power flow on line $ e \in {E^a}$, as measured at node $i$ at time $t \in T$}
\nomenclature[D,07]{$ s_{e,ij}^t$}{apparent power flow on line $ e \in {E^a}$, as measured at node $i$ at time $t \in T$}
\nomenclature[D,08]{$V_i^t$}{induced dc voltage magnitude at bus $i \in {N^d}$ at time $t \in T$}
\nomenclature[D,09]{$ z_{e} $}{1 if line {$e \in E^a$} is switched on; 0 otherwise}
\nomenclature[D,10]{$\delta_e^t$}{top oil temperature at transformer edge $e \in E^\Delta \cup E^y \cup E^\infty$ at time $t \in T$}
\nomenclature[D,11]{$\delta_{eu}^t$}{steady-state top oil temperature at transformer edge $e \in E^\Delta \cup E^y \cup E^\infty$ at time $t \in T$}
\nomenclature[D,12]{$\eta_e^t$}{hot-spot temperature at transformer edge $e \in E^\Delta \cup E^y \cup E^\infty$ at time $t \in T$}
\nomenclature[D,13]{$\theta_i^t$}{phase angle at bus $i \in{ N^a }$ at time $t \in T$}

\begin{document}

\title{\huge Analyzing and Mitigating the Impacts of GMD and EMP Events on the Electrical Grid with PowerModelsGMD.jl}

\author{
    Adam~Mate $^{1}$\orcidicon{0000-0002-5628-6509},
    Arthur~K.~Barnes $^{1}$\orcidicon{0000-0001-9718-3197},
    Russell~W.~Bent $^{1}$\orcidicon{0000-0002-7300-151X},
    and Eduardo~Cotilla-Sanchez $^{2}$\orcidicon{0000-0002-3964-3260}

\thanks{$^{1}$ The authors are with the Advanced Network Science Initiative at Los Alamos National Laboratory (LANL ANSI), Los Alamos, NM 87544 USA. Email:\{amate, abarnes, rbent\}@lanl.gov.}

\thanks{$^{2}$ The author is with the School of Electrical Engineering and Computer Science, Oregon State University, Corvallis, OR 97331 USA. Email:\{ecs\}@oregonstate.edu.}

\thanks{LA-UR-19-29623. Approved for public release; distribution is unlimited.}

}

\maketitle


\begin{abstract}
Geomagnetic disturbances and E3 high-altitude electromagnetic pulse events pose a substantial threat to the electrical grid by adversely impacting and damaging high-voltage transmission networks and equipment.
To evaluate the risks and mitigate the potential effects of these hazards, this work proposes PowerModelsGMD.jl (abbr.~PMsGMD).
PMsGMD is an open-source Julia package that solves for quasi-dc line flow and ac power flow problems in a system subjected to geomagnetically induced currents.
Unlike commercially available software solutions, it is extensible and applicable to a variety of problems. The flexibility of this framework is demonstrated by applying it to the problem of identifying mitigation strategies via line switching for a time-extended transformer heating problem.

An overview of PMsGMD is presented in this paper: introduction to its design, validation of its implementation, demonstration of its performance and effectiveness, and a description of how it may be applied to aid system-operation decisions.
\end{abstract}


\section{Introduction} \label{sec:introduction}

High-impact low-frequency (abbr.~HILF) events are the greatest threat to the continuous and reliable operation of our energy infrastructures. Such events have the potential to cause unpredictable system-wide disruptions and long-term damage in the electrical grid, a key component of the critical infrastructures.
Geomagnetic disturbances (abbr.~GMDs) and high-altitude electromagnetic pulse (abbr.~HEMP) events are among these extreme hazards \cite{introduction_1, introduction_2}.

GMDs are caused by intense solar activity: charged and magnetized particles are blown away from the Sun during severe space weather, which then interact with and disrupt the Earth's magnetic field causing rapid changes in its configuration.
These disturbances are mainly driven by large solar flares and associated coronal mass ejections during solar maximums, and by co-rotating interaction regions (high-speed solar winds) during solar minimums \cite{introduction_16, introduction_15, introduction_3}.

HEMP events are caused by nuclear explosions detonated high up in the atmosphere.
These are series of electromagnetic waveforms, covering times from nanoseconds to hundreds of seconds, that propagate to the Earth's surface.
Three main waveforms are generated during a detonation, among which the E3 late-time waveform produces electric fields with comparable time scales and area coverage as those of geomagnetic storms.
Even though GMDs tend to have higher energy, E3~HEMP events generate high enough peak field levels that makes them comparable to severe GMD events in terms of impact and caused damage \cite{introduction_4, introduction_15}.

Both GMD and E3~HEMP events pose a risk to the electrical grid by generating geomagnetically induced currents (abbr.~GICs), quasi-dc currents that appear in the conductive infrastructure and flow into the high-voltage network through the neutrals of power transformers \cite{introduction_3, introduction_4, introduction_5, introduction_15}.
GICs may adversely impact transmission networks and equipment as they have the potential to induce harmonics by causing half-cycle saturation in transformers. Harmonics may lead to the misoperation of protective devices, causing tripping of over-current relays. Premature aging, lasting damage, or complete failure of large high-voltage transformers due to overheating and thermal degradation is also a great threat. Increased reactive power consumption, caused by the circulating GICs in the system, may lead to the loss of reactive power support and to voltage collapses. In the worst case, widespread infrastructure damage and tripping of transmission lines may result in cascading failures and extended power disruptions \cite{introduction_6, introduction_7, introduction_8, introduction_9, introduction_10}.

Understanding the danger that power systems face is critically important.
This paper presents an overview of PowerModelsGMD.jl\footnote{\url{https://github.com/lanl-ansi/PowerModelsGMD.jl}} (abbr.~PMsGMD), a free and open-source package for power system simulation, which was specifically designed to evaluate the risks and mitigate the impacts of the above mentioned hazards.
The PMsGMD framework is implemented in Julia \cite{introduction_11}, a high-level just-in-time compiled programming language designed specifically for scientific computing. It is an extension to the PowerModels.jl (abbr. PMs) platform \cite{introduction_12}, a Julia/JuMP package for solving and evaluating steady-state power network optimization problems. JuMP \cite{prob_form_1} is a package for mathematical programming in Julia that specifies problems with algebraic constraints.

\vspace{0.1in}
The key contributions of this paper include:
\begin{itemize}
\item An extensible, open-source software for modeling GICs in power systems.
\item The first time-extended model of GIC that includes detrimental effects on transformers and that is appropriate for use in a mitigation optimization setting.
\item A detailed discussion on modeling requirements for GIC calculations.
\end{itemize}

\vspace{0.1in}
The remainder of this paper is organized as follows:
A review of the PMsGMD package and background of key problem formulations are provided in Section~\ref{sec:prob-formulations}; it gives context to this work and explains design goals.
Section~\ref{sec:ss-formulations} presents the implemented steady-state formulations, and Section~\ref{sec:acots-specification} describes the specifications of the time-extended GIC mitigation problem.
Section~\ref{sec:case-study} and \ref{sec:results} validates the implementations and provides a practical example of how this framework can be used both in planning and operations environments.
Section~\ref{sec:conclusion} finishes with a few concluding remarks.

\section{Problem Formulations} \label{sec:prob-formulations}

\subsection{Context of PMsGMD}

The Julia programming language \cite{introduction_11} is a high-level, high-performance, flexible dynamic language, which is appropriate for technical computing, with performance comparable to traditional statically-typed languages.
Julia has many free and open-source packages available to provide specific toolkits and capabilities for diverse applications.

PMs \cite{introduction_12} is a package for power system simulation: it provides a flexible platform for implementing and solving a wide range of steady-state network optimization and analysis problems; among others, it includes implementations of power flow (abbr.~PF), optimal power flow (abbr.~OPF), and optimal transmission switching (abbr.~OTS) problem specifications.
It relies on JuMP \cite{prob_form_1}, which provides an ideal modeling layer for the wide range of optimization problems that arise in power systems research. The use of at least one solver is required; e.g., Ipopt \cite{prob_form_2a} a fast and scalable solver for non-convex non-linear optimization. Commercial solvers such as Gurobi \cite{prob_form_2b} and CPLEX \cite{prob_form_2c} are supported as well.

In recent years PMs has emerged and filled a gap between ad-hoc research code and commercial tools such as PSS\textsuperscript{\textregistered}E and PowerWorld\textsuperscript{\textregistered} Simulator.
It is similar to the open-source MATPOWER package \cite{prob_form_3} that is widely used in the academic environment to conduct research; however, dependence on the MATLAB\textsuperscript{\textregistered} environment, licensing challenges for parallel calculations in a computing cluster, and difficulty in creating extensions present great limitations.
There have been efforts to improve GIC modeling and include related analysis in power system simulation \cite{prob_form_4, prob_form_5, prob_form_6, prob_form_7, prob_form_8}. A number of free and commercially available software solutions exist -- MATGMD \cite{prob_form_9}, PowerWorld\textsuperscript{\textregistered}'s GIC add-on \cite{prob_form_10}, and PSS\textsuperscript{\textregistered}E's GIC module \cite{prob_form_11} -- however, these all are focused on modeling and analysis with often unverifiable and non-customizable capabilities.
The PMsGMD package presented here builds on the PMs platform and provides an accessible and easy-to-handle framework to both analyze and mitigate the impacts of GMD and E3~HEMP events on electrical grids.

PMsGMD solves for quasi-dc line flow and ac power flow on a network subjected to GICs. It solves for mitigation strategies by treating the transformer overheating problem as an optimal transmission switching problem.
Due to its open-source nature, it is easy to verify and customize its operation in order to best fit the application circumstances. Due to its speed and reliability, it is suitable to be a key component of toolkits (such as \cite{introduction_17}) that monitor GMD manifestations in real-time, that predict GICs on the electrical grid, that assess risk, that enhance grid resilience by providing aid to system-operators, and that recommend modifications in the network configuration.
Consequently, PMsGMD is equally useful for both research and industry application.

\subsection{Input File Format}

PMsGMD uses several extensions to the PMs data format \cite{introduction_12} to provide input for its problem formulations.
For generality, it uses a separate dc network defined by \textit{gmd\_bus} and \textit{gmd\_branch} tables.
To correctly calculate the increased reactive power consumption of each transformer, the \textit{branch\_gmd} table adds all winding configuration related data; furthermore, the \textit{branch\_thermal} table adds thermal data necessary to determine the temperature changes in transformers.
The \textit{bus\_gmd} table includes the latitude and longitude of buses in the ac network for use in distributionally robust optimization \cite{introduction_13} or for convenience in plotting the system.

\vspace{0.1in}
The description of B4GIC \cite{introduction_14}, an included four-bus test case is presented below to demonstrate the use of the PMsGMD data format and introduce each input field. Some fields and descriptions are abbreviated due to space constraints; transformer is abbreviated as xfmr.

\vspace{0.1in}
\textit{1) GMD Bus Data Table:} Table~\ref{gmd_bus}.
\vspace{0.05in}

\begin{tabular}{rl}
\emph{parent}:& index of corresponding ac network bus \\
\emph{status}:& binary value that defines the status of bus \\ 
\emph{g\_gnd}:& admittance to ground [S] \\
\emph{name}:& a descriptive name for the bus
\end{tabular}

\begin{table}[!htbp]
\renewcommand{\arraystretch}{1.1}
\centering
\caption{mpc.gmd\_bus}
\label{gmd_bus}
\begin{tabular}{|c|c|c|c|}
\hline
\textit{parent} & \textit{status} & \textit{g\_gnd} & \textit{name} \\
\hline
\hline
1 & 1 & 5 & `dc\_sub1' \\
\hline
2 & 1 & 5 & `dc\_sub2' \\
\hline
1 & 1 & 0 & `dc\_bus1' \\
\hline
2 & 1 & 0 & `dc\_bus2' \\
\hline
3 & 1 & 0 & `dc\_bus3' \\
\hline
4 & 1 & 0 & `dc\_bus4' \\
\hline
\end{tabular}
\end{table}

\vspace{0.2in}
\textit{2) GMD Branch Data Table:} Table~\ref{gmd_branch}.
\vspace{0.05in}

\begin{tabular}{rl}
\emph{f\_bus}:& `from' bus in the gmd bus table \\
\emph{t\_bus}:& `to' bus in the gmd bus table \\
\emph{index}:& index of corresponding ac network branch \\
\emph{status}:& binary value that defines the status of branch \\
\emph{br\_r}:& branch resistance [$\Omega$] \\
\emph{br\_v}:& induced quasi-dc voltage [V] \\
\emph{len\_km}:& length of branch [km] \\
\emph{name}:& a descriptive name for the branch
\end{tabular}

\vspace{0.2in}
\textit{3) Branch GMD Data Table:} Table~\ref{branch_gmd}.
\vspace{0.05in}

\begin{tabular}{rl}
\emph{hi\_bus}: & index of high-side ac network bus \\
\emph{lo\_bus}: & index of low-side ac network bus \\
\emph{gmd\_br\_hi}: & index of gmd branch corresponding to \\ & high-side winding (two-winding xfmrs) \\
\emph{gmd\_br\_lo}: & index of gmd branch corresponding to \\& low-side winding (two-winding xfmrs) \\
\emph{gmd\_k}: & scaling factor to calculate reactive \\&  power consumption as a function\\&  of effective winding current [p.u.] \\
\end{tabular}


\begin{table*}[!htbp]
\renewcommand{\arraystretch}{1.1}
\centering
\caption{mpc.gmd\_branch}
\label{gmd_branch}
\begin{tabular}{|c|c|c|c|c|c|c|c|}
\hline
\textit{f\_bus} & \textit{t\_bus} & \textit{parent} & \textit{status} & \textit{br\_r} & \textit{br\_v} & \textit{len\_km} & \textit{name} \\
\hline
\hline
3 & 1 & 1 & 1 & 0.1 & 0 & 0 & `dc\_xf1\_hi' \\
\hline
3 & 4 & 2 & 1 & 1.001 & 170.788 & 170.788 & `dc\_br1' \\
\hline
4 & 2 & 3 & 1 & 0.1 & 0 & 0 & `dc\_xf2\_hi' \\
\hline
\end{tabular}
\end{table*}

\begin{table*}[!htbp]
\renewcommand{\arraystretch}{1.1}
\centering
\caption{mpc.branch\_gmd}
\label{branch_gmd}
\begin{tabular}{|c|c|c|c|c|c|c|c|c|c|c|}
\hline
\textit{hi\_bus} & \textit{lo\_bus} & \textit{gmd\_br\_hi} & \textit{gmd\_br\_lo} & \textit{gmd\_k} & \textit{gmd\_br\_se} & \textit{gmd\_br\_co} & \textit{baseMVA} & \textit{dispatch} & \textit{type} & \textit{config} \\
\hline
\hline
1 & 3 & 1 & -1 & 1.793 & -1 & -1 & 100 & 1 & `xfmr' & `gwye-delta' \\
\hline
1 & 2 & -1 & -1 & -1 & -1 & -1 & -1 & 1 & `line' & `none' \\
\hline
2 & 4 & 3 & -1 & 1.793 & -1 & -1 & 100 & 1 & `xfmr' & `gwye-delta' \\
\hline
\end{tabular}
\end{table*}

\begin{table*}[!htbp]
\renewcommand{\arraystretch}{1.1}
\centering
\caption{mpc.branch\_thermal}
\label{branch_thermal}
\begin{tabular}{|c|c|c|c|c|c|c|c|c|c|}
\hline
\textit{xfmr} & \textit{temp\_amb} & \textit{hs\_inst\_lim} & \textit{hs\_avg\_lim} & \textit{hs\_rated} & \textit{to\_time\_c} & \textit{to\_rated} & \textit{to\_init} & \textit{to\_inited} & \textit{hs\_coeff} \\
\hline
\hline
1 & 25 & 280 & 240 & 150 & 71 & 75 & 0 & 1 & 0.63 \\
\hline
0 & -1 & -1 & -1 & -1 & -1 & -1 & -1 & -1 & -1 \\
\hline
1 & 25 & 280 & 240 & 150 & 71 & 75 & 0 & 1 & 0.63 \\
\hline
\end{tabular}
\end{table*}


\begin{tabular}{rl}
\emph{gmd\_br\_se}: & index of gmd branch corresponding to \\& series winding (auto-xfmrs) \\
\emph{gmd\_br\_co}: & index of gmd branch corresponding to \\& common winding (auto-xfmrs) \\
\emph{baseMVA}: & [MVA] base of xfmr \\
\emph{type}: & type of branch -- `xfmr' / `transformer' \\& or `line' or `series\_cap' \\
\emph{config}: & winding configuration of transformer \\
\end{tabular}

\vspace{0.25in}
\textit{4) Branch Thermal Data Table:} Table~\ref{branch_thermal}.
\vspace{0.05in}

\begin{tabular}{rl}
\emph{xfmr}: & binary value that defines if branch is an \\& 'xfmr' or not \\
\emph{temp\_amb}: & ambient temperature of xfmr [$\degree\mathrm{C}$] \\
\emph{hs\_inst\_lim}: & 1-hour hot-spot temp. limit of xfmr [$\degree\mathrm{C}$]\\
\emph{hs\_avg\_lim}: & 8-hour hot-spot temp. limit of xfmr [$\degree\mathrm{C}$]\\
\emph{hs\_rated}: & hot-spot temperature-rise of xfmr \\& at rated power [$\degree\mathrm{C}$]\\
\emph{to\_time\_c}: & top-oil temperature-rise time-constant \\& of xfmr [min]\\
\emph{to\_rated}: & top-oil temperature-rise of xfmr at \\& rated power [$\degree\mathrm{C}$]\\
\emph{to\_init}: & initial top-oil temperature of xfmr [$\degree\mathrm{C}$]\\
\emph{to\_inited}: & binary value that defines the initial top-\\&oil temperature of xfmr: 1 with \textit{to\_init} \\& value, 0 with steady-state value \\
\emph{hs\_coeff}: & relationship of hot-spot temperature rise \\& to  Ieff [$\degree\mathrm{C/A}$]
\end{tabular}

\vspace{0.25in}
\textit{5) Bus GMD Data Table:} Table~\ref{bus_gmd}.
\vspace{0.05in}

\begin{tabular}{rl}
\textit{lat}: & latitude coordinate of ac network bus and \\& corresponding dc network bus \\
\textit{lon}: & longitude coordinate of ac network bus and \\& corresponding dc network bus \\
\end{tabular}

\begin{table}[!htbp]
\renewcommand{\arraystretch}{1.1}
\centering
\caption{mpc.bus\_gmd}
\label{bus_gmd}
\begin{tabular}{|c|c|}
\hline
\textit{lat} & \textit{lon} \\
\hline
\hline
40 & -89 \\
\hline
40 & -87 \\
\hline
40 & -89 \\
\hline
40 & -87 \\
\hline
\end{tabular}
\end{table}

\newpage
The related fields of Branch GMD Data (Table~\ref{branch_gmd}) and Branch Thermal Data (Table~\ref{branch_thermal}) tables are set to ($-1$) when a system component is not a transformer.
PMsGMD is able to model both fixed (spatially uniform) and realistic (non-uniform) electric fields as it does not model line coupling.

\subsection{Modeling GIC Impact}

The goal of modeling is to accurately simulate GICs and determine the exact threat at any particular place and time in a power system.
As the level of detail required to model a system is more than what is needed in a traditional positive-sequence simulation, it is a complex task \cite{prob_form_6}. GICs are dependent on system characteristics (geographical location of substations, resistance of system components, characteristics of transformers), geomagnetic source fields (amplitude, frequency content, spatial characteristics), and the Earth conductivity structure (modeling method, substation grounding resistance, influence on geo-electric fields); all these need to be considered in the modeling process.
The greatest challenge is the availability of data; verifiable information on the threat and detailed description of the system is needed for accurate results.

First, the equivalent dc network must be created from the ac power system \cite{introduction_3}.
As the impedances of each phase of the system are identical -- i.e., the phase conductors (transmission lines and transformer windings in each phase) provide identical parallel paths for GIC flows -- the calculations of GICs are performed for a single phase only, and the same result applies to each phase.
Each resistive branch is replaced by its corresponding admittance value and voltage sources. Transmission lines with series capacitive compensation are omitted in the dc model as series capacitors block the flow of GIC. Transformers are modeled with their winding resistance to the substation neutral and, in the case of auto-transformers, both series and common windings are represented explicitly.

\vspace{0.05in}
PMsGMD currently supports ``\textit{gwye-gwye}'', ``\textit{gwye-delta}'', ``\textit{delta-delta}'', and ``\textit{gwye-gwye-auto}'' transformer winding configurations, where \textit{gwye} stands for grounded-wye.
When unknown, this can be estimated: \\
$\bullet$ generator step-up (abbr.~GSU) transformers: \textit{delta-gwye} with delta on low side \\
$\bullet$ load transformers: \textit{delta-gywe} with delta on high side \\
$\bullet$ \textit{gwye-gywe-auto}-transformers: connect portions of the transformer grid where voltage ratio $\leq$ 3 \\
$\bullet$ \textit{gwye-gwye} transformers: connect portions of the transformer grid where voltage ratio $>$ 3.

\vspace{0.05in}
In practice, actual geo-electric fields vary with geographical locations. Using a common assumption that the North and East components of the geo-electric field are constant in the geographical area of the transmission line \cite{prob_form_13, prob_form_14, prob_form_6}, the induced voltage $\mathcal{V}$ is calculated as
\begin{equation}
\label{eq:voltage source}
\allowdisplaybreaks
\small
\mathcal{V}=\mathcal{E}_NL_N + \mathcal{E}_EL_E = |\mathcal{E}|(\sin(\phi)L_N + \cos(\phi) L_E)
\end{equation}
where $L_N, L_E, \mathcal{E}_N, \mathcal{E}_E,$ and $\phi$ are described in Appendix~I of \cite{prob_form_6}.
Due to short length, $\mathcal{V} = 0$ is assumed for transformers.

This formulation assumes that $\phi$ is measured in counter-clockwise direction from the positive x axis; in geography, it is convention to measure $\phi$ in clockwise direction from the positive y axis, consequently conversion is necessary sometimes.
The substation grounding resistance is often approximated by using nonlinear regression, based on substation size, maximum kV level, and number of incoming lines as predictor variables.

\subsection{Transformer Modeling}

The formulation presented in \cite{prob_form_16} contains two equivalent power system models: one for computing the ac power flows and one for computing the GICs; for some formulations, coupling constraints are added to link GICs to the ac power flow model in the form of reactive power losses. The main difference between them occurs in transformer modeling. 
AC power flow models typically model transformers as a single edge with a voltage transformation (phase shift and tap change). GIC models, on the other hand, require models of transformers that include details of the series and common windings, as well as other transformer components \cite{prob_form_15}.

\begin{figure*}[!htbp]
\centering
\subfloat[\textit{gywe-delta}]{\includegraphics[width=1.05in]{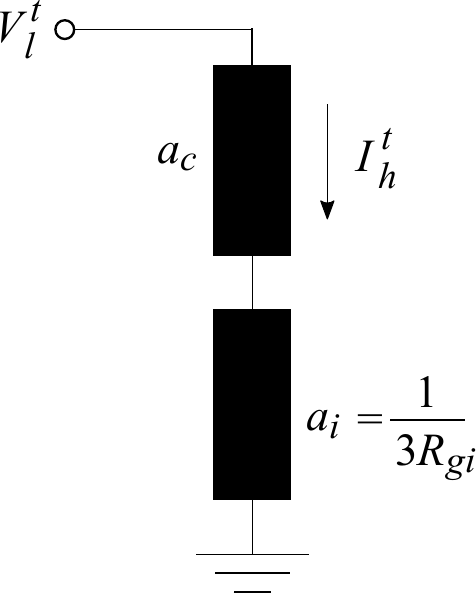}
\label{fig:gywe-delta-dc-equiv}}
\hfil
\subfloat[\textit{gywe-gywe}]{\includegraphics[width=1.45in]{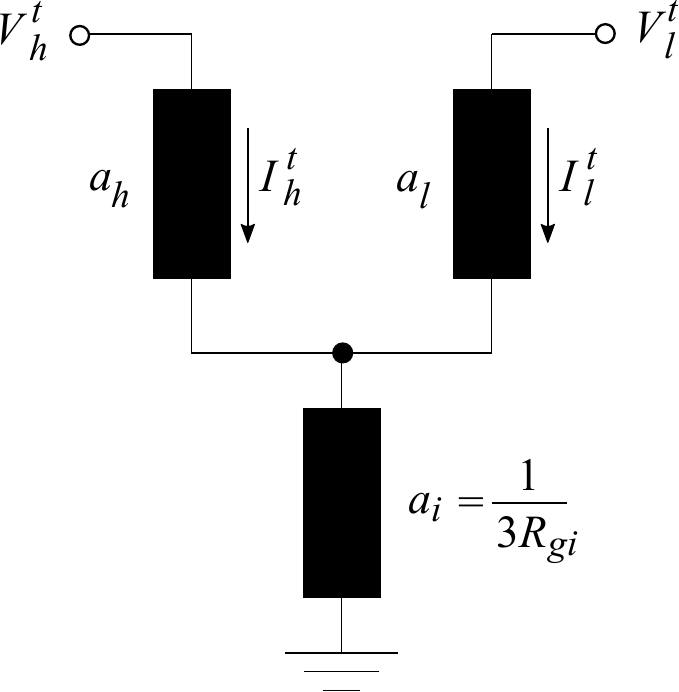}
\label{fig:gywe-gywe-dc-equiv}}
\hfil
\subfloat[\textit{gywe-gywe-auto}]{\includegraphics[width=0.9in]{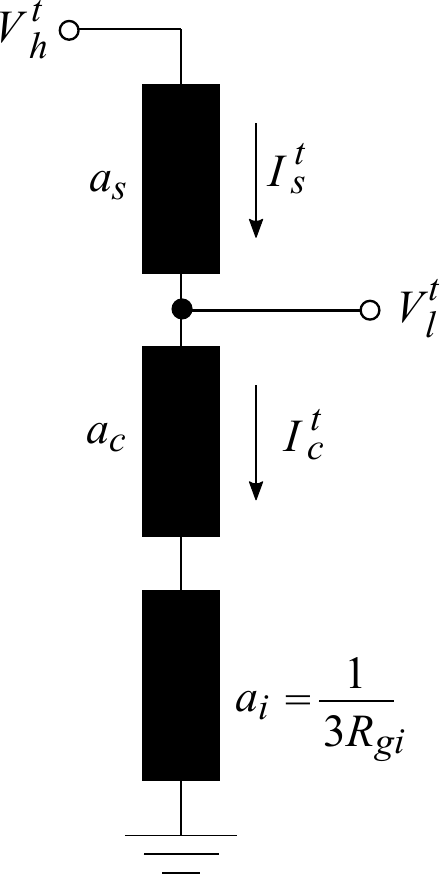}
\label{fig:gywe-gwye-auto-dc-equiv}}
\caption{DC equivalent circuit of different transformer-windings}
\label{fig:dc_eqv_circuits}
\end{figure*}

Effective GIC in the ac network is computed using the following set of equations: \\
\noindent for $ \forall e \in E^a, t \in T $
\begin{equation}
\label{eq:effective_gic}
\allowdisplaybreaks
\small
\widetilde{I}_{e}^t =
\begin{cases}
\abs{I^t_{e^H}} & \text{if $e \in E^\Delta$} \vspace{.1cm} \\
\abs{\frac{\alpha_e I^t_{e^H} + I^t_{e^L} }{\alpha_e}  } & \text{if $e \in E^y$} \vspace{.1cm}  \\
\abs{\frac{\alpha_e I^t_{e^S} + I^t_{e^C} }{\alpha_e + 1}  } & \text{if $e \in E^\infty$} \vspace{.1cm} \\
0 & \text{otherwise}
\end{cases}
\end{equation}
\noindent where Case~1 models a \textit{gywe-delta} GSU transformer (dc equivalent circuit for this case is illustrated in Fig.~\ref{fig:gywe-delta-dc-equiv}), Case~2 models a \textit{gywe-gywe} transformer (dc equivalent circuit for this case is illustrated in Fig.~\ref{fig:gywe-gywe-dc-equiv}), and Case~3 models a \textit{gywe-gwye-auto}-transformer (dc equivalent circuit for this case is illustrated in Fig.~\ref{fig:gywe-gwye-auto-dc-equiv}).

\vspace{0.05in}
As \cite{prob_form_16} notes, most test networks in the literature neglect GSU transformers.
GSUs are used to connect the output terminals of generators to the transmission network. The GSUs and the neutral leg ground points they provide are critical when modeling GICs and methods to mitigate the impact of GICs. Consequently, in network models that lack GSU transformers, the model is assumed to include GSU transformers using the method discussed in \cite{prob_form_16}. 
Here it is assumed that GSU transformers are \textit{delta-gywe} transformers.

\vspace{0.05in}
In this paper the model of \cite{prob_form_16} is adapted to a multi-period dispatch to model the GIC impacts on power systems.
The contribution is a model of temperature change over time as a function of changing GIC and ac flows. Instead of the equivalent Norton representation, the simpler Thevenin representation is used as PMs \cite{introduction_12} supports series voltage sources.
The continuous-time formulation of the transformer hot-spot thermal model is a first order linear dynamic model that is defined in terms of hot-spot temperature rise over top-oil temperature.

Given the active power $p_{e,ij}^t$ [p.u.] and reactive power $q_{e,ij}^t$ [p.u.] flowing into the input terminal of the transformer at time $t$ [min], the apparent power is defined by
\begin{equation}
\label{xfmr-thermal_apparent}
\allowdisplaybreaks
\small
s_{e,ij}^t = \sqrt{(p_{e,ij}^t)^2 + (q_{e,ij}^t)^2}
\end{equation}

\noindent Fractional loading of the transformer is defined by
\begin{equation}
\label{xfmr-thermal_loading}
\allowdisplaybreaks
\small
k_e^t = \frac{s_{e,ij}^t}{\overline{s}_{e}}
\end{equation}

\noindent Steady-state top-oil temperature rise $\delta$ over ambient is then
\begin{equation}
\label{xfmr-thermal_to-temp-ss}
\allowdisplaybreaks
\small
\delta_{eu}^t = \delta_e^r k^2
\end{equation}
\noindent where $\delta_e^r$ is the top-oil temperature rise in [$\degree\mathrm{C}$] at rated apparent power. 
For a typical transformer, $\delta_e^r = 75[{\degree}\mathrm{C}]$ \cite{prob_form_18}.

\vspace{0.05in}
From \cite{prob_form_19}, the top-oil temperature rise over ambient is
\begin{equation}
\label{xfmr-thermal_to-temp-amb}
\allowdisplaybreaks
\small
\delta_e(t) = \delta_{eu} (1 - e^{t/\tau_e})
\end{equation}
\noindent where $\tau_e$ is the time constant of the top-oil temperature [min]. A typical value is 71 [mins] \cite{prob_form_18}.
By using a bilinear transform, the hot-spot model is reformulated as a difference equation that is suitable for optimization and simulation \cite{prob_form_12}:
\begin{equation} \label{ac_temp}
\delta_{e}^t = \frac{1}{1 + \zeta_e}(\delta_{eu}^t + \delta_{eu}^{t-1}) - \frac{1-\zeta_e}{1+\zeta_e}\delta_e^{t-1}
\end{equation}
\noindent where $t$ is the sample index,  $\zeta_e = 2\tau_e/\Delta$ is the discretized time constant, and  $\Delta$ is the sampling period.

The time constants of the hot-spot temperature rise over the top-oil temperature are typically negligible compared to the top-oil temperature dynamics.
The absolute value of the transformer hot-spot is \\
\noindent for $ \forall e \in E^\Delta \cup E^y \cup E^\infty, t \in T $
\begin{equation}
\label{eq:xfmr-hs-abs} \rho_e^t + \delta_e^t + \eta_e^t
\end{equation}
\noindent where $\rho_e^t$ is the ambient temperature, and assumed to be $\rho_e^t= 25 \enskip [\degree\mathrm{C}]$, and $\eta_e^t$ is the hot-spot temperature rise:
\begin{equation}
\label{eq:xfmr-hs-rise}
\allowdisplaybreaks
\small
\eta_e^t = R_e \widetilde{I}_{e}^t
\end{equation}
\noindent where $R_e$ is assumed to be $R_e = 0.63\enskip [\degree\mathrm{C}/A]$ and $\widetilde{I}_{e}^t$ is the transformer effective GIC; i.e. the hot-spot temperature rise is linear with respect to the effective GIC.

\section{Steady-State Formulations} \label{sec:ss-formulations}

Before presenting the time-extended formulation, first the common industry and academic steady-state formulations implemented in PMsGMD are discussed.

\vspace{0.1in}
\noindent \textit{1) GIC DC:} quasi-dc power flow. \\
Solves steady-state dc currents on lines resulting from induced dc voltages on lines.

\vspace{0.1in}
\noindent \textit{2) GIC $\xrightarrow{}$ AC -- OPF:} sequential quasi-dc power flow and ac optimal power flow. \\
Solves for quasi-dc voltages and currents, and uses the calculated quasi-dc currents through transformer windings as inputs to an AC-OPF in order to calculate the increase in transformer reactive power consumption.

\vspace{0.1in}
\noindent \textit{3) GIC + AC -- OPF:} ac optimal power flow coupled with a quasi-dc power flow. \\
Solves for quasi-dc voltages and currents, and solves the AC-OPF concurrently. The dc network couples to the ac network by means of reactive power loss in transformers.
This formulation does not model increases in transformer reactive power consumption caused by changes in the ac terminal voltages. Additionally, it may report higher reactive power consumption than reality as it relaxes the ``effective'' transformer quasi-dc winding current magnitude.

\vspace{0.1in}
\noindent \textit{4) GIC + AC -- MLS:} ac minimum-load-shed coupled with a quasi-dc power flow. \\
Solves the minimum-load shedding problem for a network subjected to GIC with fixed topology. It uses load shedding to protect the system from GIC-induced voltage collapse and transformer over-heating.
 
\vspace{0.1in}
\noindent \textit{5) GIC + AC -- OTS:} ac optimal transmission switching with load shed coupled with a quasi-dc power flow. \\
Solves the minimum-load shedding problem for a network subjected to GIC, where lines and transformers can be opened or closed. It uses transmission-switching to protect the system from GIC-induced voltage collapse and transformer over-heating.

\section{Time-series Extension of the \\ AC-OTS Problem Specification} \label{sec:acots-specification}

Below the problem specification for the AC-OTS use case of PMsGMD is described.
This problem uses the top-oil temperature dynamics described in Section~\ref{sec:prob-formulations} to link time-periods. Line switching configurations are shared across all time-periods. 
In this paper the dc power flow approximation of the full ac non-convex power flow constraints is used to ease computational burden. However, switching to the full set of ac power flow constraints or relaxations is seamless given the PMs framework. 

\vspace{0.05in}
\subsubsection{Objective function} \hfill \\
\begin{equation}
\label{eq:dc-ots_obj}
\allowdisplaybreaks
\small
\min \smashoperator{\sum_{g \in G, t \in T}} \kappa_g^0 + \kappa_{g}^1 p_g^t + \kappa_{g}^2 (p_g^t)^2 \end{equation}

\vspace{0.05in}
\subsubsection{DC power flow approximation equations} \hfill \\

\vspace{-0.05in}
for $ \forall i\in {N^a}, t \in {T} $
\begin{equation}
\label{eq:dc-ots_powflow-approx_1}
\allowdisplaybreaks
\small
\smashoperator{\sum_{e \in {E_i^+}} } p_{e,ij}^t - \smashoperator{\sum_{e \in {E_i^-}} } p_{e,ij}^t
= \sum_{g \in G_i} p_g^t - p_{i}^t - g_i
\end{equation}

\vspace{0.05in}
for $ \forall e \in {E^a}, t \in {T} $
\begin{subequations}
\label{eq:dc-ots_powflow-approx_2}
\allowdisplaybreaks
\small
\begin{align}
p_{e,ij}^t=z_{e} b_{e}(\theta_i^t-\theta_j^t) 
\label{dc-ots_pij}
\\
p_{e,ji}^t=-p_{e,ij}^t
\label{dc-ots_pji}
\end{align}
\end{subequations}

\vspace{0.05in}
\subsubsection{Operational limit constraints} \hfill \\

\vspace{-0.05in}
for $ \forall e \in {E^a}, t \in {T} $
\begin{subequations}
\label{eq:dc-ots_op-lim-const_1}
\allowdisplaybreaks
\small
\begin{align}
p_{e,ij}^t \leq z_{e}\overline{s}_{e}
\label{eq:dc-ots_op-lim-const_1a}
\\
p_{e,ji}^t \leq z_{e}\overline{s}_{e}
\label{eq:dc-ots_op-lim-const_1b}
\\
|\theta_i^t-\theta_j^t| \leq z_{e}\overline{\theta} + (1-z_{e})\theta^M
\label{eq:dc-ots_op-lim-const_1c}
\end{align}
\end{subequations}

\vspace{0.05in}
for $ \forall g \in G, t \in {T} $
\begin{equation}
\label{eq:dc-ots_op-lim-const_2}
\allowdisplaybreaks
\small
\underline{p}_g \leq p_g^t \leq \overline{p}_g
\end{equation}

\vspace{0.05in}
\subsubsection{GIC effects on transformers} \hfill \\

\vspace{-0.05in}
for $ \forall i \in {N^d} $
\begin{equation}
\label{eq:gic_effect_1_gic}
\allowdisplaybreaks
\small
\smashoperator{ \sum_{e \in E^+_i}}I_e^t - \smashoperator{\sum_{e \in E^-_i}} I_e^t = a_{i}V_i^d
\end{equation}

\vspace{0.05in}
for $ \forall e \in \mathcal{E}^{d}, t \in {T} $
\begin{equation}
\label{eq:gic_effect_2_Id}
\allowdisplaybreaks
\small
I_{e}^t=z_{\overrightarrow{e}}a_{e}(V_i^t-V_j^t + \mathcal{V}_e^t)
\end{equation}

\vspace{0.05in}
for $ \forall e \in {E^a}, t \in {T} $
\begin{equation}
\label{eq:gic_effect_3_Idmag}
\allowdisplaybreaks
\small
\widetilde {I}_{e}^t \geq +/-
\begin{cases}
{I^t_{e^H}} & \text{if $e \in E^\Delta$} \vspace{.1cm} \\
{\frac{\alpha_e I^t_{e^H} + I^t_{e^L} }{\alpha_e}  } & \text{if $e \in E^y$} \vspace{.1cm} \\
{\frac{\alpha_e I^t_{e^S} + I^t_{e^C} }{\alpha_e + 1}  } & \text{if $e \in E^\infty$} \vspace{.1cm} \\
0 & \text{otherwise}
\end{cases}
\end{equation}

\vspace{0.05in}
for $ \forall e \in \mathcal{E}^{\tau} $
\begin{equation}
\label{eq:gic_effect_4_Idub}
\allowdisplaybreaks
\small
0 \leq {\widetilde{I}_{e}^d }\leq \overline{I}_{e}^d
\end{equation}

\vspace{0.05in}
\begin{equation}
\label{eq:gic_effect_5_heating}
\allowdisplaybreaks
\small
\mathrm{Eq.} \;\; \eqref{ac_temp}, \eqref{eq:xfmr-hs-rise}
\end{equation}

\vspace{0.05in}
for $ \forall e \in E^\Delta \cup E^y \cup E^\infty, t \in T $
\begin{equation}
\label{eq:gic_effect_6_xfmr-hs-limit}
\allowdisplaybreaks
\small
\rho_e^t + \delta_e^t + \eta_e^t \le \mathcal{T}_e
\end{equation}

\vspace{0.05in}
for $ \forall e \in {\mathcal{E}^a}, g \in G $
\begin{equation}
\label{eq:gic_effect_7_binary}
\allowdisplaybreaks
\small
z_{e} \in \{0, 1 \}
\end{equation}

\vspace{0.05in}
\subsubsection{Supporting constraints} \hfill \\

\vspace{-0.05in}
for $ \forall e \in E^a, t \in T $
\begin{equation}
\label{eq:support_const_gen-effective_gic_help}
\allowdisplaybreaks
\small
\widetilde{I}_{e}^t \le z_e \overline{I}_{e}^d
\end{equation}

\vspace{0.1in}
The objective function \eqref{eq:dc-ots_obj} minimizes total generator dispatch costs and load shedding costs.
Constraints \eqref{eq:dc-ots_powflow-approx_1}-\eqref{eq:dc-ots_powflow-approx_2} describe the dc power flow physics and engineering constraints associated with a power system.
Constraint \eqref{eq:dc-ots_powflow-approx_1} models power balance at each node (Kirchoff's Law). 
Constraints \eqref{dc-ots_pij} and \eqref{dc-ots_pji} model the dc approximation of ac power flow on each transmission line with on-off variables {$z_{e}$} (Ohm's Law)..

Constraints \eqref{eq:dc-ots_op-lim-const_1}-\eqref{eq:dc-ots_op-lim-const_2} describe the operational limits of the power system.
Constraints \eqref{eq:dc-ots_op-lim-const_1a} and \eqref{eq:dc-ots_op-lim-const_1b} model the thermal limits of lines in both directions.
Constraint \eqref{eq:dc-ots_op-lim-const_1c} applies bounds on the phase angle difference between two buses. Constraint \eqref{eq:dc-ots_op-lim-const_2} models the capacity of power generation.
The dc circuit and effects associated with the GMD are formulated in constraints \eqref{eq:gic_effect_1_gic}-\eqref{eq:gic_effect_7_binary}.
An edge ($e \in \mathcal{E}^d$) in the dc circuit is linked to an edge in the ac circuit ($\overrightarrow{e}$).
Constraints \eqref{eq:gic_effect_1_gic} and \eqref{eq:gic_effect_2_Id} formulate the GIC flow on each dc line by applying Kirchhoff's current law using the Thevenin representation.

The GIC on a line is determined by the induced current source and the quasi-dc voltage difference between two buses \cite{prob_form_6}. 
GIC flow is forced to 0 by $z_{\overrightarrow{e}}$ when $\overrightarrow{e}$ is switched off.
Constraint \eqref{eq:gic_effect_3_Idmag} computes the effective (non-negative) GIC on each ac edge by relaxing the absolute values in \eqref{eq:effective_gic} with two inequalities. The effective GIC is calculated at the time scale of the ac circuit.
Constraint \eqref{eq:gic_effect_4_Idub} models the maximum allowed value of GIC flowing through a transformer for safe operation.
Constraint \eqref{eq:gic_effect_5_heating} computes the ac and GIC hot-spot heating temperature at every time step using Equations \eqref{ac_temp} and \eqref{eq:xfmr-hs-rise}.
Constraint \eqref{eq:gic_effect_6_xfmr-hs-limit} forces transformer temperatures below a thermal limit.
Finally, constraint \eqref{eq:support_const_gen-effective_gic_help} explicitly forces the effective GIC to 0 when an edge is inactive.

\vspace{0.1in}
Code Block 1 (on Page~\pageref{CodeBlock1}) shows how this specification is implemented in PMsGMD in Julia. Some function names are abbreviated due to space constaints.

\section{Case Study} \label{sec:case-study}

\begin{figure}[!htbp]
\centering
\includegraphics[width=0.475\textwidth]{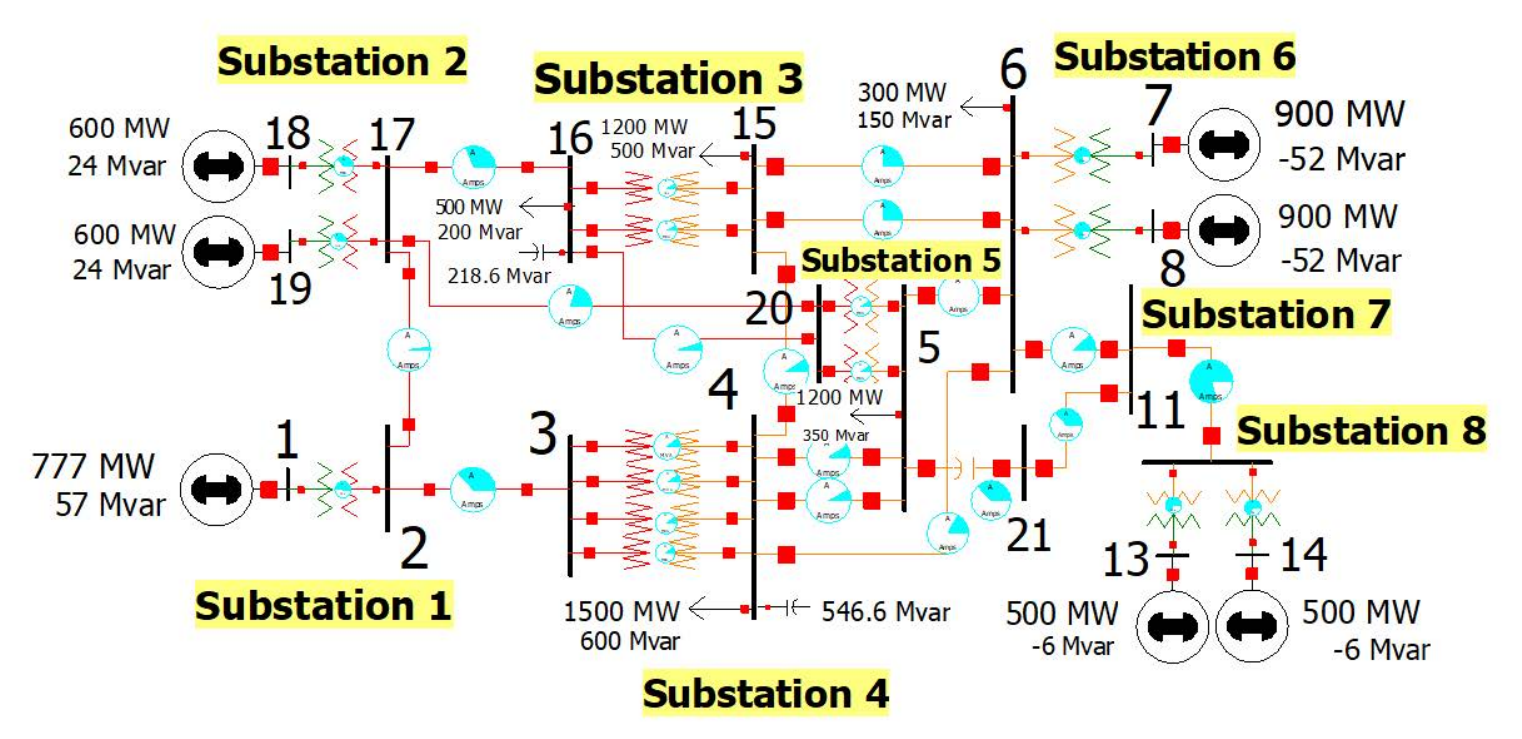}
\caption{One-line diagram of the used case study system}
\label{fig:epri21-oneline}
\end{figure}

\begin{figure}[!htbp]
\centering
\fbox{\includegraphics[width=0.35\textwidth]{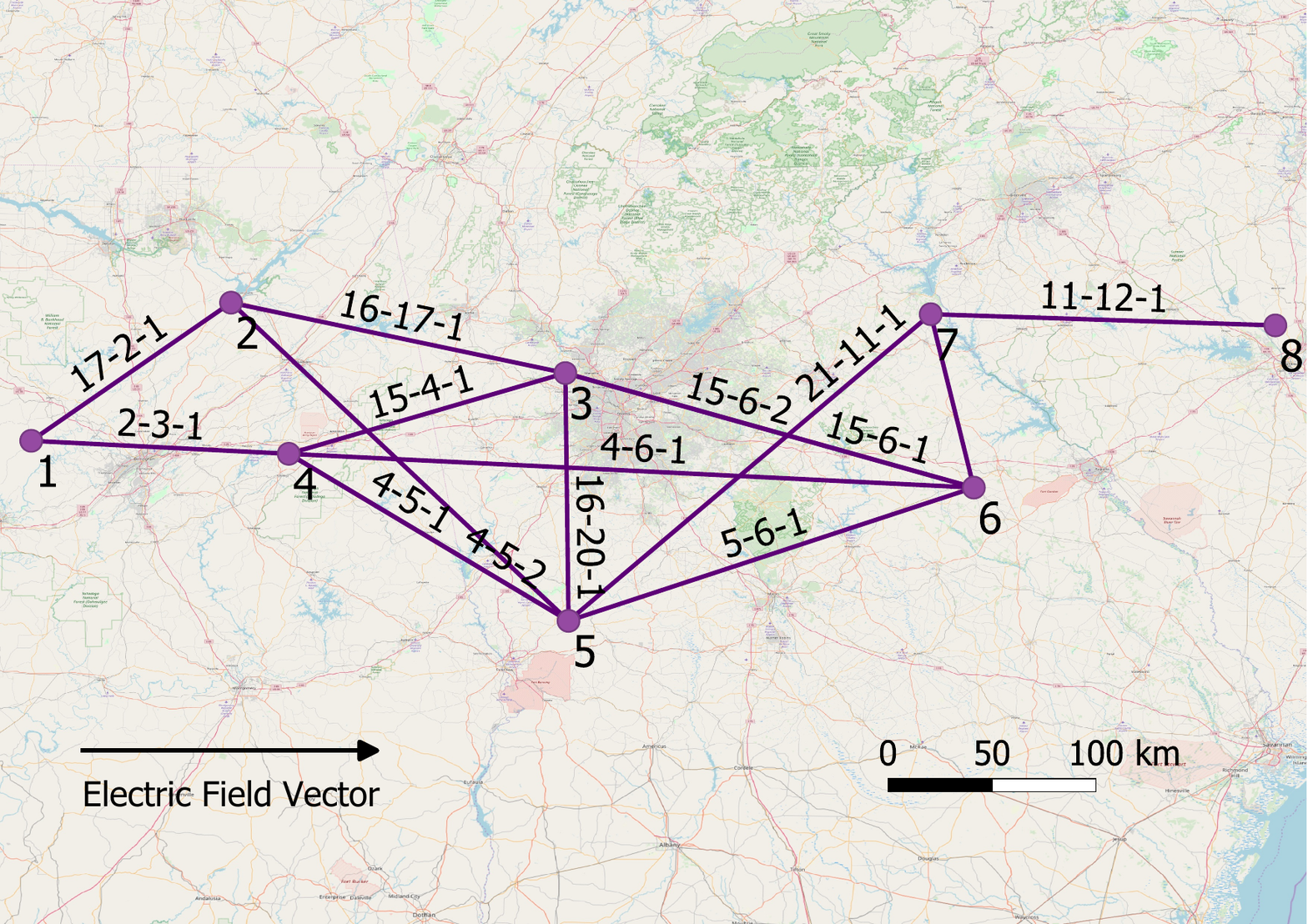}}
\caption{Map of the used case study system and electric field vector}
\label{fig:epri21-map}
\end{figure}

To demonstrate the application and performance of PMsGMD, in addition validate the implementation of the AC-OTS ``ac optimal transmission switching with load shed coupled with a quasi-dc power flow'' problem specification (presented in Section~\ref{sec:acots-specification}), a GMD case study is carried out.
The hypothetical network designed by Horton et al.~\cite{prob_form_13} for GIC modeling software validation is used. This 21-bus extra high voltage power system model -- depicted in Figs.~\ref{fig:epri21-oneline} and \ref{fig:epri21-map} -- consists of 345~[kV] and 500~[kV] lines and transformers, and is centered over the State of Tennessee, USA. The model includes single transmission lines as well as some that occupy the same transmission corridor; the substations feature both conventional and auto-transformers, furthermore, series and neutral connected GIC blocking devices are also included.

\vspace{0.05in}

\vspace{0.05in}

\begin{figure}[!htbp]
\centering
\includegraphics[width=0.485\textwidth]{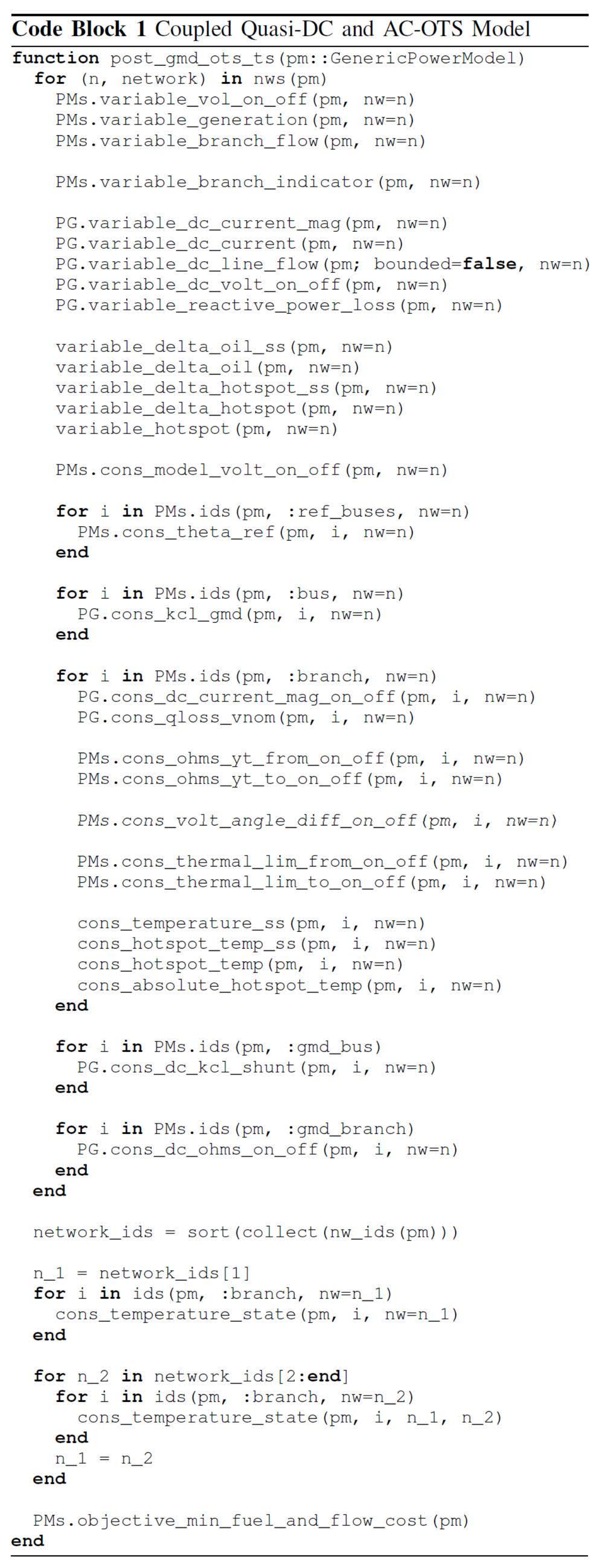}
\label{CodeBlock1}
\end{figure}

The Horton case study system provides all parameters necessary for calculations of GICs, including bus-substation connectivity and substation geographical locations.
In order to simulate a realistic system operation scenario, a number of lines are disconnected (as indicated in the above figures) assuming planned system-maintenance and unforeseen outages due to a GMD event; consequently, the size of the network is reduced to 11 buses.
Table~\ref{tab:br-status} summarizes these initial assumptions for each branch.

\begin{table}[!htbp]
\centering
\caption{Branch Statuses in the Case Study System}
\begin{tabular}{|r|r|c|r|c|c|r|r|}
        \hline
        \textit{i} & \textit{j} & \textit{Ckt.} & \textit{Type} & $z_e^{nom}$ & $z_e$ & $p_{e,ij}$ & $I_e$ \\
        \hline \hline
 1 &  2 & 1 &         xf & 1  &       1 &   8.3 &   0.0 \\
 2 &  3 & 1 &       line & 1  &       1 &   8.3 &   0.0 \\
 3 &  4 & 1 &         xf & 1  &       1 &   2.1 & -56.6 \\
 3 &  4 & 2 &         xf & 1  &       1 &   2.1 & -56.6 \\
 3 &  4 & 3 &         xf & 1  &       1 &   2.1 & 127.8 \\
 3 &  4 & 4 &         xf & 1  &       1 &   2.1 & 127.8 \\
 4 &  5 & 1 &       line & 1  &       1 &  -6.7 & 368.8 \\
 4 &  5 & 2 &       line & 1  & {\bf 0} &   0.0 &   0.0 \\
 4 &  6 & 1 &       line & 1  & {\bf 0} &   0.0 &   0.0 \\
 5 &  6 & 1 &       line & 1  &       1 & -18.7 & 368.8 \\
 5 & 20 & 1 &         xf & 0  &       0 &   0.0 &   0.0 \\
 5 & 20 & 2 &         xf & 0  &       0 &   0.0 &   0.0 \\
 5 & 21 & 1 & series\_cap & 0  &       0 &   0.0 &   0.0 \\
 6 &  7 & 1 &         xf & 1  &       1 &  -8.3 &  21.5 \\
 6 &  8 & 1 &         xf & 1  &       1 &  -8.3 &  21.5 \\
 6 & 11 & 1 &       line & 1  &       1 &  -5.0 & 325.9 \\
11 & 12 & 1 &       line & 1  &       1 &  -5.0 & 325.9 \\
12 & 13 & 1 &         xf & 1  &       1 &  -5.0 & 325.9 \\
12 & 14 & 1 &         xf & 0  &       0 &   0.0 &   0.0 \\
15 &  4 & 1 &       line & 0  &       0 &   0.0 &   0.0 \\
15 &  6 & 1 &       line & 0  &       0 &   0.0 &   0.0 \\
15 &  6 & 2 &       line & 0  &       0 &   0.0 &   0.0 \\
16 & 15 & 1 &         xf & 0  &       0 &   0.0 &   0.0 \\
16 & 15 & 2 &         xf & 0  &       0 &   0.0 &   0.0 \\
16 & 17 & 1 &       line & 0  &       0 &   0.0 &   0.0 \\
16 & 20 & 1 &       line & 0  &       0 &   0.0 &   0.0 \\
17 &  2 & 1 &       line & 0  &       0 &   0.0 &   0.0 \\
17 & 18 & 1 &         xf & 0  &       0 &   0.0 &   0.0 \\
17 & 19 & 1 &         xf & 0  &       0 &   0.0 &   0.0 \\
17 & 20 & 1 &       line & 0  &       0 &   0.0 &   0.0 \\
21 & 11 & 1 &       line & 0  &       0 &   0.0 &   0.0 \\
\hline
    \end{tabular}
    \label{tab:br-status}
\end{table}

\begin{figure}[!htbp]
\centering
\includegraphics[width=0.425\textwidth]{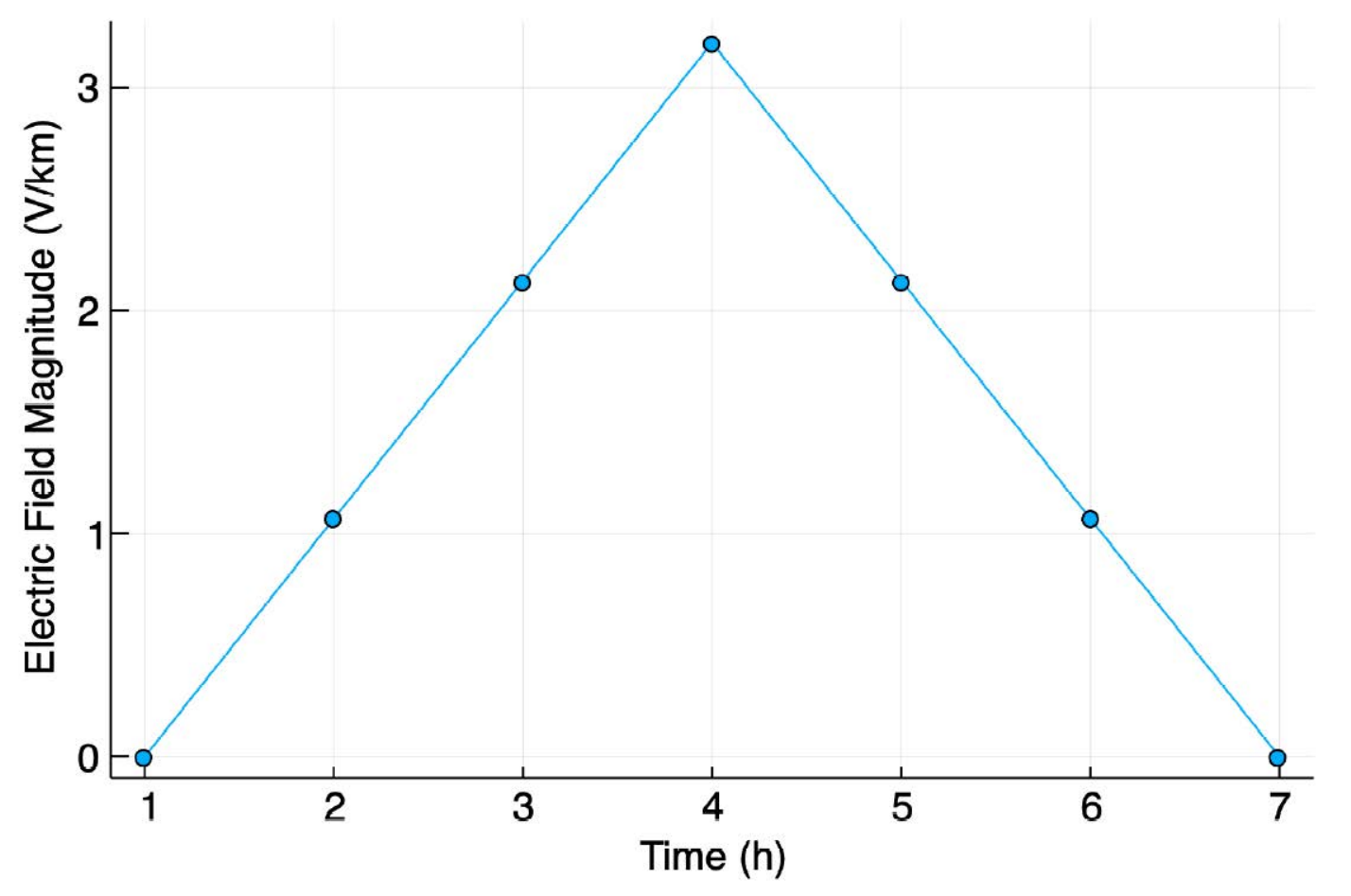}
\caption{Time variation of the electric field magnitude}
\label{fig:e-field-mag}
\end{figure}

The system is subjected to a time-varying uniform East-West electric field (as indicated in Fig.~\ref{fig:epri21-map}), which varies linearly from 0 $\mathrm{V/km}$ to 3.2 $\mathrm{V/km}$ over the first 3 hours and then back down linearly to 0 $\mathrm{V/km}$ over the remaining 3 hours of the simulation, illustrated in Fig.~\ref{fig:e-field-mag}.

\section{Evaluation of Results} \label{sec:results}

The simulation was performed on a computer with a quad-core Inter\textsuperscript{\textregistered} Core\textsuperscript{TM} i5-6500 3.2GHz CPU, and 8GB RAM. The used software was Julia v1.0 \cite{introduction_11}, using Gurobi v8.1 \cite{prob_form_2b} for MacOS as nonlinear solver.

Parts of the results are presented in Table~\ref{tab:br-status}: the columns $z_e^{nom}$ and $z_e$ list the nominal branch statuses along with the branch statuses produced by the optimizer, respectively.
The mitigation strategy determination takes 21~[sec]. The optimizer chooses to relieve GIC flowing through the transformers at Substations 1, 4 and 6; it accomplishes this by opening the longest predominantly East-West directional line \textit{4-6-1}, along with one of the lines in the \textit{4-5} transmission corridor.

\begin{figure}[!htbp]
\centering
\includegraphics[width=0.425\textwidth]{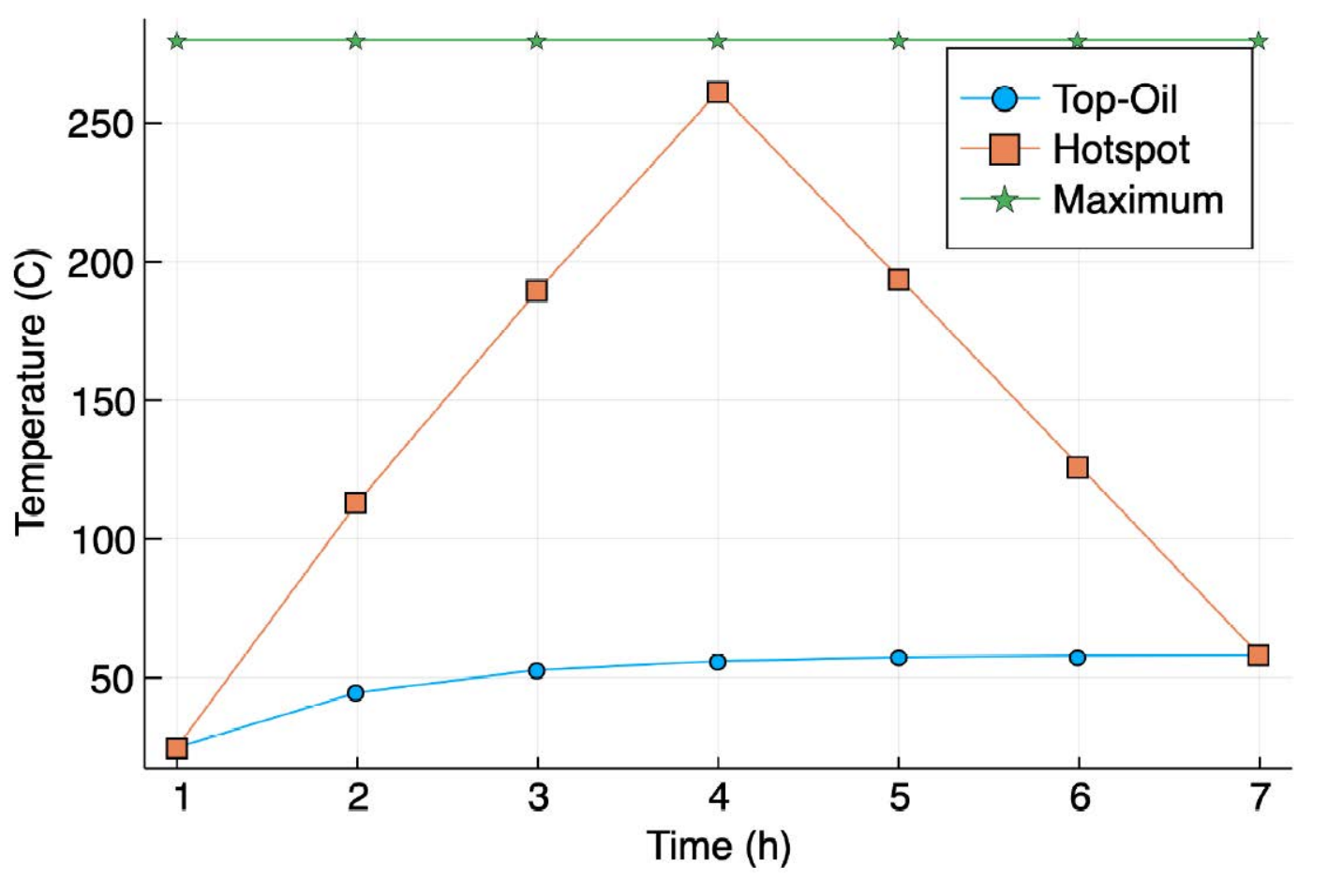}
\caption{Hot-spot temperature for transformer 12-13-1}
\label{fig:br-temp}
\end{figure}

Fig.~\ref{fig:br-temp} illustrates the temperature of transformer \textit{12-13-1}, located at substation 8 of Fig.~\ref{fig:epri21-map}; the designed mitigation strategy kept its temperature below the instantaneous limit of $280\degree\mathrm{C}$.
This admittedly high instantaneous limit is selected with a bias to supplying load power at the cost of significant loss-of-life to transformer insulation; nevertheless, this limit can be adjusted as needed.

\section{Conclusion} \label{sec:conclusion}

As the threat of GMD and E3~HEMP events continues to pose a substantial risk to our energy infrastructures, accurate simulation and effective mitigation becomes increasingly important.
This paper presented an extensible open-source modeling framework for both analyzing and optimally mitigating the impacts of such hazards.

By design, PMsGMD is suitable to monitor disturbance-manifestations in real-time and predict GICs on the electrical grid.
In addition, by extending PMsGMD with the problem of mitigating transformer heating by switching lines given a time-varying electric field, along with techniques to improve the scalability of the problem, a GIC mitigation capability is enabled; providing mitigation-strategies in a reasonable amount of time demonstrates the utility of this modular framework.

The performed case study validates implementation and demonstrates performance; further simulations on larger size networks are future tasks.
Other extensions to the presented GIC power flow problem are possible under the PMsGMD framework, such as optimal GIC blocker placement, harmonic load flow, and scenario-based problem formulations.

\newpage
\vspace{0.1in}
\noindent
\textbf{Acknowledgements} This work was supported by the U.S. DOE LDRD Program at Los Alamos National Laboratory under the \emph{``Impacts of Extreme Space Weather Events on Power Grid Infrastructure: Physics-Based Modelling of Geomagnetically-Induced Currents (GICs) During Carrington-Class Geomagnetic Storms"} project.


\bibliographystyle{unsrt}
\bibliography{arxiv.bib}



\begin{figure}[!htbp]
\centering
\includegraphics[width=0.48\textwidth]{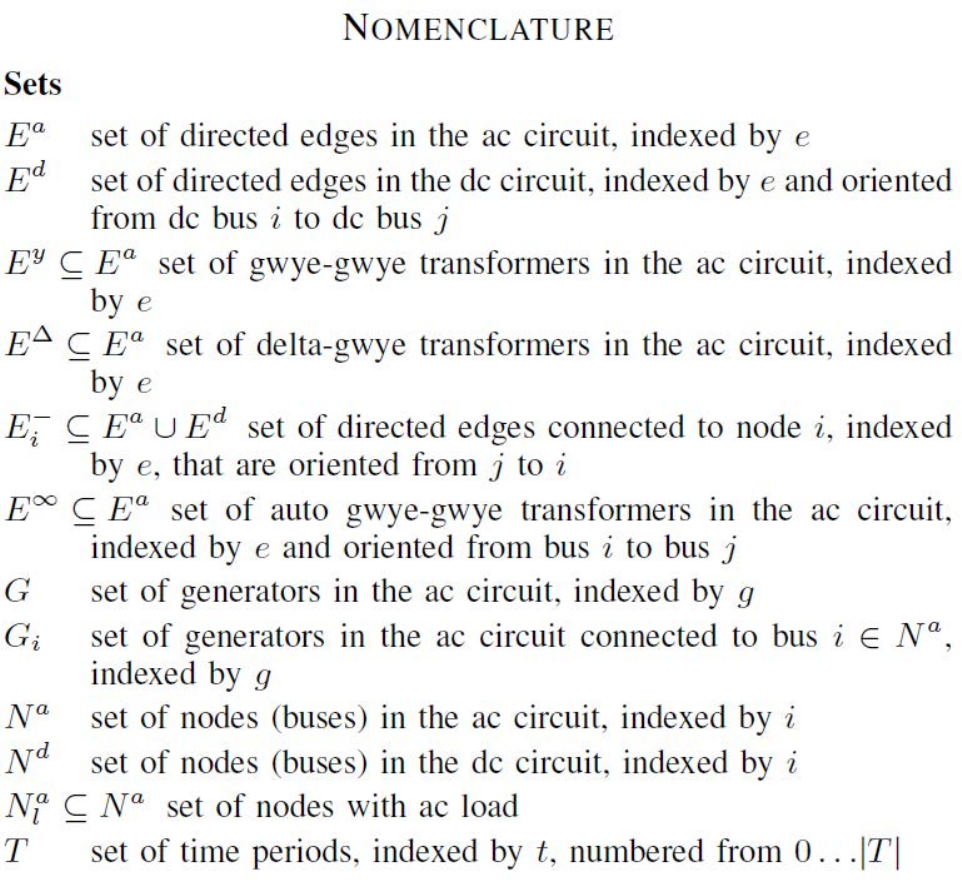}
\end{figure}

\begin{figure}[!t]
\centering
\includegraphics[width=0.485\textwidth]{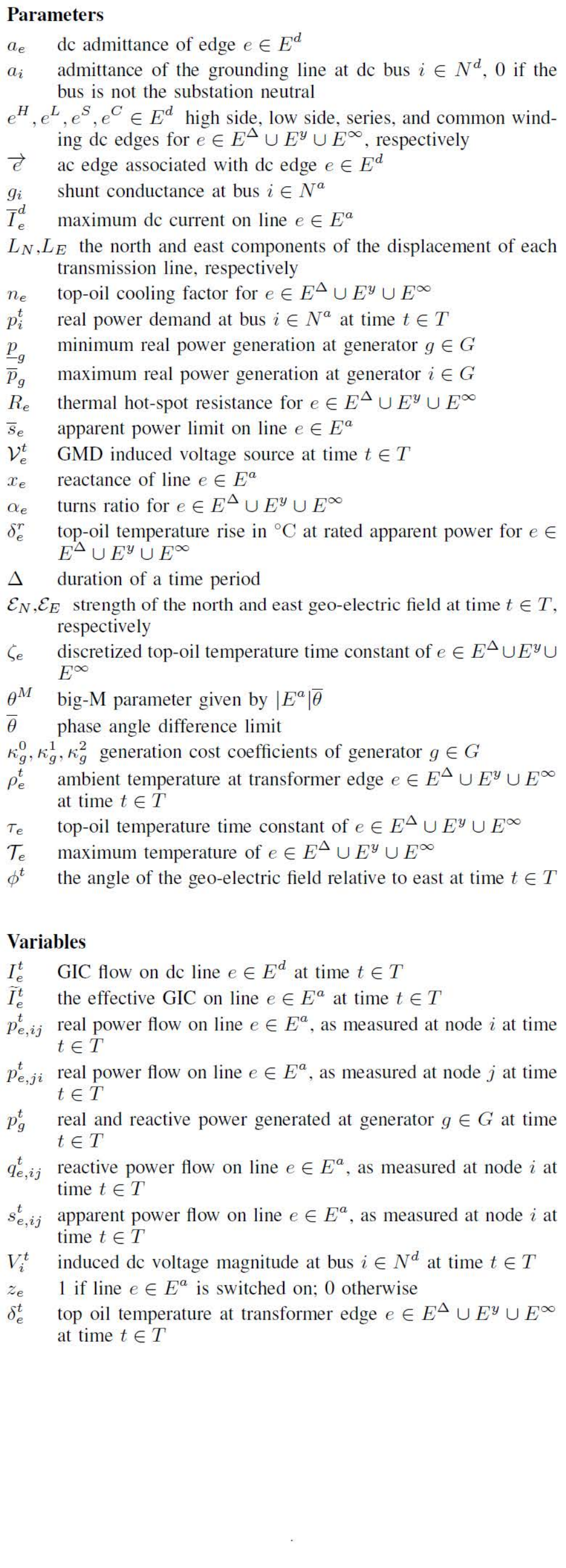}
\end{figure}

\end{document}